# Dissipative soliton generation and real-time dynamics in microresonator-filtered fiber lasers

Mingming Nie[1, 3, *], Bowen Li[1, 3], Kunpeng Jia[2], Yijun Xie[1], Jingjie Yan[1], Shining Zhu[2], Zhenda Xie[2], and Shu-Wei Huang[1, *]

[1]Department of Electrical, Computer and Energy Engineering, University of Colorado Boulder, Boulder, Colorado 80309, USA
[2]National Laboratory of Solid State Microstructures, Nanjing University, Nanjing 210093, China
[3]These authors contribute equally
[*]Corresponding author: mingming.nie@colorado.edu, shuwei.huang@colorado.edu

## Abstract

Optical frequency combs in microresonators (microcombs) have a wide range of applications in science and technology, due to its compact size and access to considerably larger comb spacing. Despite recent successes, the problems of self-starting, high mode efficiency as well as high output power have not been fully addressed for conventional soliton microcombs. Recent demonstration of laser cavity soliton microcombs by nesting a microresonator into a fiber cavity, shows great potential to solve the problems. Here we comprehensively study the dissipative soliton generation and interaction dynamics in a microresonator-filtered fiber laser in both theory and experiment. We first bring theoretical insight into the mode-locking principle, discuss the parameters effect on soliton properties and provide experimental guidelines for broadband soliton generation. We predict chirped bright dissipative soliton with flat-top spectral envelope in microresonators with normal dispersion, which is fundamentally infeasible for externally driven case. Furthermore, we experimentally achieve soliton microcombs with large bandwidth of ~10 nm and high mode efficiency of 90.7%. Finally, by taking advantage of an ultrahigh-speed time magnifier, we study the real-time soliton formation and interaction dynamics and experimentally observe soliton Newton's cradle. Our study will benefit the design of the novel, high-efficiency and self-starting microcombs for real-world applications.

## Introduction

Optical frequency combs (OFCs) have been the cornerstones and key enabling technologies for many scientific breakthroughs in precision frequency metrology [1,2], ultra-stable time keeping [3,4], coherent comb spectroscopy [5,6] and other fields [7–10]. Enormous progress has been made towards conventional OFCs based on self-starting mode-locked lasers, where efficient laser gain provides the comb energy. On the other hand, the emerging dissipative Kerr soliton (DKS) microcombs [11–14] have recently served as a promising complement due to its compact size and access to considerably larger comb spacing. However, fundamentally limited by external drive configuration and low coupling rate of the microresonators with high quality (Q) factor, the achieved DKS microcombs exhibit low output power and low mode efficiency theoretically limited to 5% for bright DKS [15]. Therefore, it is challenging to directly apply the microcombs for applications demanding high comb power [16,17]. Besides, the bi-stability and strong thermal nonlinearity prohibits the self-starting of DKS microcombs and requires complex active electronics for soliton initiation and stabilization [12,18,19]. Despite the efforts of nonlinear self-injection locking [20,21], cascaded Brillouin-Kerr process [22], synchronous pumping [23], dark-soliton combs [24] and impedance matching [25], self-starting microcombs with high output power and high mode efficiency are still challenging.

Recent pioneering work [26] reports soliton microcombs in a microresonator-filtered fiber laser [27], which provides the unique opportunity to realize laser cavity soliton (LCS) microcombs with advantages of self-starting, theoretical high mode efficiency up to 96% [26] and no need of a tunable single-frequency laser. From the perspective of external driving DKS, LCS is formed through synchronous pumping [23] where the driven pulse comes from the DKS itself in the microresonator. Despite the advantages presented in this solo case, the soliton mode-locking principle and corresponding dynamics in microresonator-filtered fiber lasers have not been studied especially in the experiment, mainly due to the challenge of real-time measurement for ultrahigh repetition rate mode-locked pulses. Fortunately, the powerful time magnifier system can help to surpass the electronic limits and study microresonator dynamics at ultrahigh repetition rates of several tens of GHz or more [28–30]. Being capable to characterize non-repetitive and arbitrary waveforms in real time with sub-picosecond temporal resolution in a single shot [31–34], the time magnifier system represents an ideal tool to study the soliton dynamics, which helps to gain insight into the underlying physics and achieve low-noise microcombs in microresonator-filtered fiber lasers.

Here we theoretically and experimentally study the dissipative soliton generation and real-time dynamics in a fiber laser nested by a high-Q fiber Fabry–Pérot (FFP) microresonator. We provide new theoretical insight into the mode-locking principle and provide experimental guidelines for stable soliton with broad bandwidth. Apart from anomalous group velocity dispersion (GVD) induced solitons, we predict chirped bright soliton with flat-top spectral envelope from normal-GVD microresonators, in contrast to the dark soliton in the externally driven system [24]. Subsequently, according to our proposed experimental guidelines, we experimentally achieve self-starting soliton microcombs with broad 3-dB bandwidth of ~10 nm and repetition rates from ~10 GHz to ~100 GHz. The soliton output power is as high as ~12 mW with mode efficiency up to 90.7%. Finally, by taking advantage of an ultrahigh-speed time magnifier, we study the real-time dynamics of soliton formation and interaction with coexisting continuous-wave (CW) components in microresonator-filtered fiber lasers. We observe soliton Newton's cradle, where a single soliton elastically collides with a soliton molecule [35–37], akin to macro-objects [38–40]. Our study will benefit the design of the novel, high-efficiency and self-starting microcombs for real-world applications.

## Results

**Mode-locking principle.** Without loss of generality, the scheme for dissipative soliton generation in microresonator-filtered fiber lasers is shown in Fig. 1a. A high-Q FP microresonator is embedded within a longer fiber cavity consisting of an Erbium doped fiber amplifier (EDFA). The FP microresonator can also be replaced by a with a ring microresonator with add-drop ports as in [26]. The final oscillating modes are the eigen-frequencies for both the fiber laser cavity and the microresonator. By properly adjusting the cavity length, only single frequency is assumed to be resonant in each resonance of the microresonator (Fig. 1b) if the free spectral range (FSR) of the fiber cavity is larger than the linewidth of the microresonator, resulting in super-mode suppression.

According to the mode-locking principle of conventional fiber lasers, a saturable absorber (SA) is essential to suppress CW lasing and select pulses from background noise [41–43]. Moreover, proper pulse shaping mechanism is required to accomplish the pulse self-reproduction [43]. Although soliton is proved to be the solution of the microresonator-filtered lasers by both simulation and experiment, it still remains veiled how the soliton outbeats the CW component and eventually stabilizes itself. Here we theoretically and numerically find that the nested microresonator acts as two roles: (1) the SA; (2) the nonlinearity booster (see theory in Supplementary Information Section S1). Furthermore, intracavity pulses can be shaped via the balance between Kerr nonlinearity and either dispersion or strong intracavity spectral filtering, depending on both GVDs of the microresonator and the EDFA.

Nested microresonators with anomalous GVD are first studied theoretically to identify their roles. The dispersion, nonlinearity and gain dispersion in the EDFA, as well as intracavity bandpass filter (BPF), are neglected in this case to clearly show the pulse initiation and stabilization are mainly accomplished by the microresonator itself. The influence of the neglected parameters will be discussed in the next section. As shown in the subfigures (i) and (ii) of Fig. 1c, for anomalous-GVD microresonators, transmission of the lower branch with red detuning exhibits obvious SA effect: higher input power induces higher transmission thus lower loss for the fiber cavity. The subfigure (iii) of Fig. 1c demonstrates LCS self-starting process from amplified spontaneous emission (ASE). The pulse is firstly selected by the SA from amplified intensity noise, reaches the threshold of modulation instability (MI) and eventually evolves into a stable soliton. In other words, red-detuned microresonator favors soliton with high peak power and high transmission instead of CW solution [44] (see Supplementary Information Section S2). The pulse shaping is accomplished by the balance between cavity-enhanced nonlinearity and anomalous GVD in the microresonator. The EDFA only provides the soliton gain without playing roles in pulse selecting and pulse shaping, which is verified by the identical soliton profiles and corresponding spectra before and after the microresonator [subfigures (iv) and (v) of Fig. 1c]. In contrast to the red-detuned case, this SA-like nonlinear transmission does not occur in the blue-detuned microresonator. Therefore, the fiber cavity length should be adjusted carefully to realize red detuning thus consequent mode-locking due to the detuning-dependent SA.

Besides, bright dissipative soliton is also feasible in this kind of lasers with inserted normal-GVD microresonators and an intracavity BPF, where the mode-locking principle is similar to all-normal-dispersion (ANDi) fiber lasers [45] and flat-top spectrum can be achieved [the subfigure (iv) of Fig. 1d]. This provides a simple method for broadband comb with bright pulses in normal-GVD microresonators, which is not feasible in the conventional case [24]. Of note, the dispersion, nonlinearity and gain dispersion in the EDFA are also neglected here.

Similarly, the red-detuned normal-GVD microresonator acts as both the SA [subfigures (i) and (ii) of Fig. 1d] and the nonlinearity booster. In the subfigures (iv) and (v) of Fig. 1d, the pulse accumulates large nonlinearity phase shifts in the normal-GVD microresonator, simultaneously expands its spectrum and enlarges the pulse width, leading to highly chirped pulse. The BPF outside the microresonator performs the pulse shaping in both frequency and time domain due to the frequency-time mapping and accomplishes the pulse self-reproduction, in the same way of ANDi lasers. Additionally, soliton can also self-start from ASE as shown in the subfigure (iii) of Fig. 1d.

The acting roles of both SA and nonlinearity booster by red-detuned microresonators, are of advantage to achieve mode-locking with ultrahigh repetition rate, which is out of reach for conventional mode-locked lasers due to: (1) conventional SAs usually have high saturation threshold, only response to pulse with high peak power and can not discriminate CW components and the low-energy pulses with ultrahigh repetition rate; (2) in conventional mode-locked lasers, low pulse energy results in insufficient Kerr nonlinearity for longitudinal mode interaction and mode-locking.

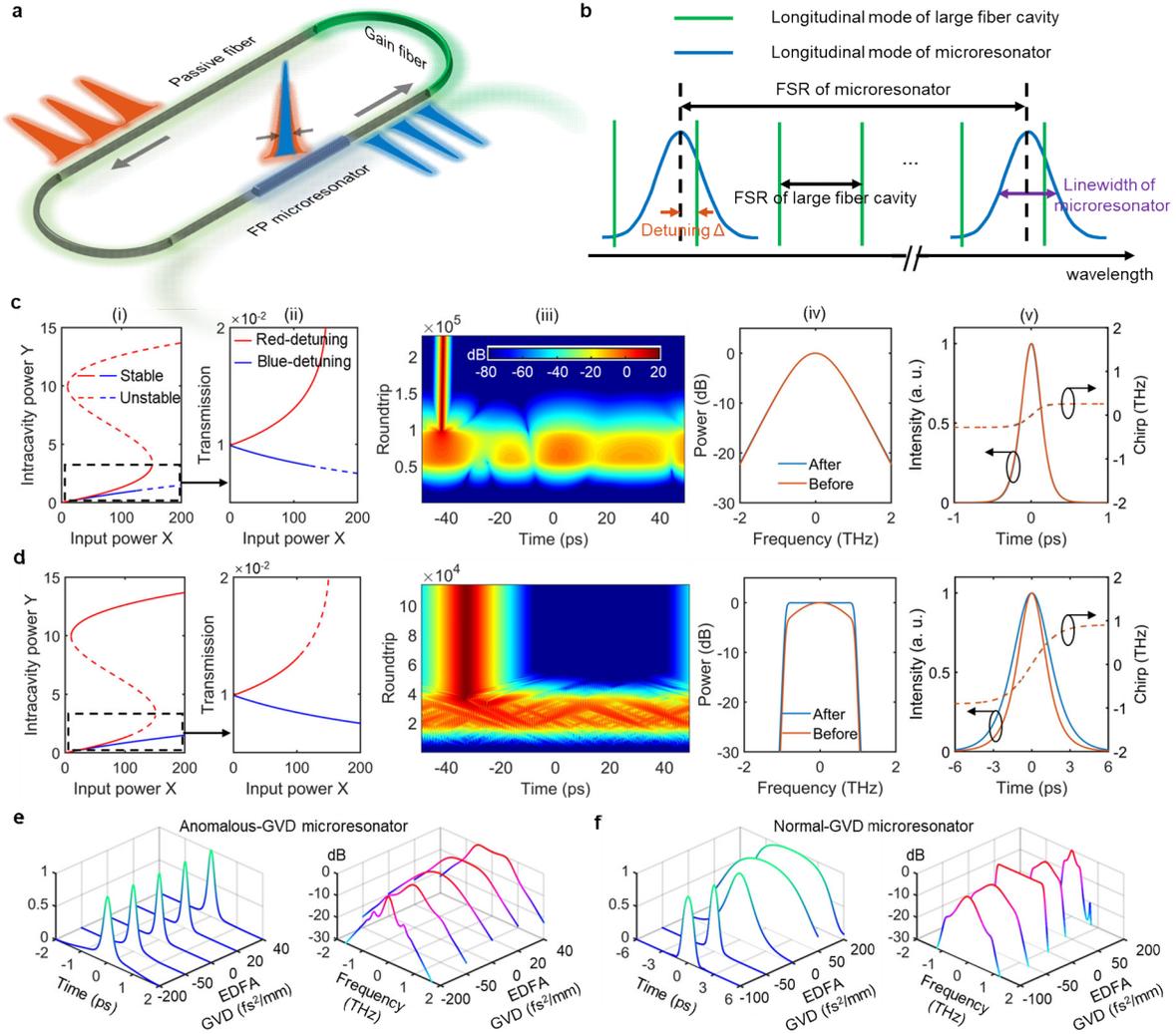

Fig. 1. Mode-locking principle and theoretical analysis. (a) Scheme for dissipative soliton generation in microresonator-filtered fiber lasers. A microresonator is nested in a fiber laser cavity. (b). Cold-cavity spectral distribution, showing the FSRs of the large fiber cavity and microresonator, resonance and corresponding linewidth of the microresonator. The detuning represents the relative position of the oscillating modes within the microresonator. In (b), red detuning is shown. (c)(d) Simulations for microresonator with anomalous GVD (c) and normal GVD (d). The nonlinearity, GVD and gain dispersion of the EDFA are neglected. Subfigure (i): bifurcation of homogeneous states. Subfigure (ii): transmission of the lower branch. Subfigure (iii): self-initiation of dissipative soliton from ASE. Subfigures (iv) and (v): spectra and pulse profiles before and after the microresonator. (e)(f) Simulated mode-locked pulse profiles (left) and spectra (right) for microresonators with anomalous GVD (e) and normal GVD (f) by varying the GVD of the EDFA.

**Experimental guidelines.** Despite the theoretical mode-locking principle, many other parameters should be taken into account in the experiment for stable dissipative soliton with broad bandwidth. Since (1) nonlinearity is negligible for the EDFA due to the ultrahigh-repetition-rate induced low pulse energy; (2) the laser gain dispersion is almost fixed, we will focus on the dispersions of both the microresonator and EDFA, as well as bandwidth of the inserted BPF.

Figures 1e and 1f demonstrate dissipative soliton generation in a fiber laser nested by a microresonator with anomalous GVD and normal GVD, respectively. By varying the EDFA's dispersion from anomalous to normal regime, mode-locked pulse width first narrows around near-zero EDFA GVD and then broadens, while the spectral bandwidth first broadens and then narrows. Therefore, broad soliton bandwidth can be achieved with near-zero EDFA GVD. In addition, similar to conventional DKS, small microresonator GVD benefits large soliton bandwidth, regardless of its sign (see Supplementary Information Section S3). Moreover, opposite signs of both microresonator and EDFA GVDs will lead to dispersive wave emission (see Supplementary Information Section S4) [46] and cause soliton destabilization, which can be solved by an intracavity narrowband BPF to achieve stable soliton.

Overall, experimental guidelines can be summarized for stable dissipative soliton with broad bandwidth. (1) It is better to employ high-Q microresonators with large nonlinear parameters to lower the threshold, narrow the comb linewidth, lower the requirement of short fiber cavity length and suppress the super-modes. (2) It is beneficial to introduce microresonators with relatively high transmission through add-drop ports to lower the insertion loss. However, compromise should be made between the high Q and high transmission through add-drop ports. (3) Broad soliton bandwidth can benefit from engineered small GVDs for both the microresonator and the EDFA. (4) Compromise between the stable soliton bandwidth and intracavity spectral filtering should be made when the microresonator and EDFA exhibit different signs of GVD (see Supplementary Information Section S5).

**Experimental generation of dissipative soliton.** The experimental setup is shown in Fig. 2a. According to the above-proposed experimental guidelines, we carefully design the microresonator-filtered fiber laser to achieve dissipative soliton with broad bandwidth. (1) Our FFP microresonator is made of highly nonlinear fiber (HNLF) with large nonlinear coefficient of 11.5 $W^{-1} \cdot km^{-1}$. High Q factor of $2 \times 10^7$ is achieved with a resonance linewidth of 9.7 MHz (Fig. 2b), which is much smaller than the FSR of the large fiber cavity (~42 MHz), resulting in the suppression of super-modes since only single frequency can exist in each resonance of the microresonator. The FSR of the FFP microresonator is 10.18 GHz which can be ready for X-band applications. (2) The cold-cavity transmission of 7% (including the coupling loss from the input fiber end-face) is relatively large, leading to a pump threshold as low as 30 mW. (3) The GVD of the FFP microresonator is ~-2.5 $fs^2/mm$ (see Materials and Methods) and the average dispersion of the EDFA is engineered to be as small as -4.4 $fs^2/mm$. (4) A BPF (Semrock, NIR01-1570/3-25) with a flat-top profile of 8.9-nm bandwidth (3 dB) is introduced in the fiber cavity to adjust the lasing center wavelength. Other details of the experimental setup in Fig. 2a can be found in Methods.

At pump power of ~600 mW, dissipative soliton can self-start with random numbers ranging from 1 to 10, corresponding to comb repetition rate ranging from ~10 to ~100 GHz (see Supplementary Information Section S6). The output power is ~12 mW, which is almost one-order-magnitude higher compared with conventional DKS, and the mode efficiency is as high as 90.7%, approaching the theoretical limit of 96% [26].

Figure 2c and Figure 3 show examples of single soliton and perfect soliton crystals with 2, 4 and 6 FSRs, respectively. Other multiple soliton states can be found in the Supplementary Information Section S7. The mode-locked states are confirmed by the clear radio frequency (RF) beat notes of the comb repetition rate (insets of Fig. 2c, Fig. 3a and Fig. 3b). In particular, we measure the single sideband (SSB) phase noise of the single soliton as shown in Fig. 2d. The SSB phase noises at 10 kHz, 100 kHz, and 1 MHz offset frequencies are -97 dBc/Hz, -117 dBc/Hz, and -137 dBc/Hz, respectively. The large phase noises below 1 kHz offset frequencies, which can be significantly reduced by active feedback loops, are attributed to the soliton relative intensity noise (RIN) resulting from the active gain fiber and the noise coupling from the environment, while the peak at 70 kHz comes from the RIN of the EDFA pump (Fig. 2e). The fundamental comb linewidth is less than 32 mHz (Fig. 2f), which is 38 dB better than the conventional external cavity diode lasers (ECDLs), due to the ultrahigh Q induced laser phase noise suppression. Even with narrower comb linewidth, the soliton phase noise is at the same level of the results in an on-chip microresonator pumped by an ECDL [47], indicating detuning noise mainly contributes to the soliton phase noise.

Moreover, the perfect soliton crystals and corresponding soliton numbers are also confirmed by the real-time pulse evolution with clean background (right panels of Figs. 3a-3c) using the time magnifier, which can be viewed as a powerful and ultrahigh-speed optical oscilloscope (see Methods and Supplementary Information Section S8 for details). The recording time window of the time magnifier is determined by the pump pulse width to ~285 ps. The peak voltage difference of the solitons along the fast time axis is due to the non-uniform responsivity inside the recording time window caused by the wavelength-dependent four-wave-mixing (FWM) efficiency (see Supplementary Information Section S8). The calibrated pulses with similar intensities [insets of Fig. 3 (right panels)] by the responsivity curve (Fig. S8d in Supplementary Information) as well as the equal temporal distances, verify the perfect soliton crystals with high-contrast and smooth spectra. Besides, owing to the imperfect synchronization between the soliton signal and the pump pulses of the time magnifier (see Supplementary Information S8), the measured soliton intensity evolution along the slow time axis exhibits intensity variation and discontinuity when the same soliton gradually walks off from the recording time window.

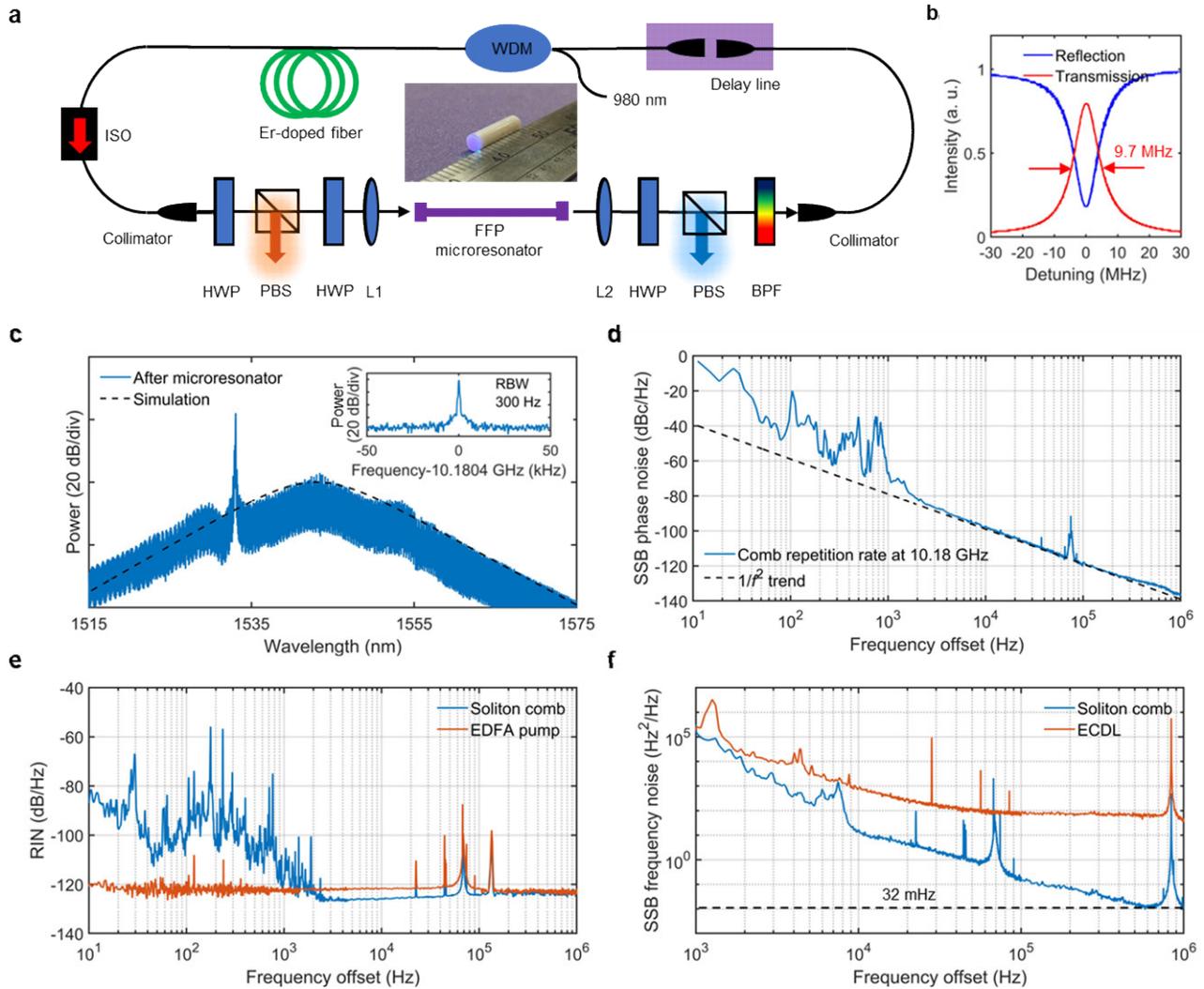

Fig. 2. Experimental generation of dissipative soliton. (a) Schematic of experimental setup. ISO: isolator, WDM: wavelength division multiplexer, HWP: half-wave plate, PBS: polarizing beam splitter, L1 and L2: lens, BPF: band-pass filter. (b) Transmission and reflection of the FFP microresonator. (c) Spectrum of the single soliton. Inset: RF beat note of the soliton repetition rate. (d) SSB phase noise spectrum of the single soliton. (e) RIN of the single soliton and the EDFA pump. (f) SSB frequency noise spectra of a comb line and a conventional ECDL (Toptica, CTL1550).

Benefiting from the careful dispersion engineering around zero dispersion, the 3-dB bandwidth of the spectra output from the microresonator is ~10 nm, which agrees excellently with the simulated results (dashed black lines) and represents three times larger than the reported result [26]. The enlarged spectrum after the FFP microresonator indicates the boosted nonlinearity induced spectrum broadening and soliton shaping inside the microresonator, which validates the roles of the microresonator according to the above-mentioned mode-locking principle. Further increase of the soliton bandwidth can result from a broadband intracavity BPF and will eventually be limited by the gain bandwidth.

Interestingly, a single comb line stands out from the soliton spectral envelope and always locates around the short-wavelength edge of the BPF (see the noise floor of the spectra before the microresonator). However, no transmission spike is found for the BPF (see Supplementary Information Section S9). We attribute the standing-out comb line with high intensity to the phase shift at the BPF edges [48–51] (see Supplementary Information Section S9). In contrast to the red-detuned soliton comb lines, the BPF induced CW peak is believed to be blue detuned [52] resulting from the phase shift. The CW peak does not belong to the soliton spectrum since it locates in the middle of two adjacent soliton comb lines. In addition, the coexisting CW component can cause soliton long-range interaction with each other and lead to perfect soliton crystal generation in Fig. 3 [53–58], without the help of avoid mode crossings since our FFP microresonator is single-mode. The interleaved weak-intensity comb lines (peaks indicated by orange solid circles in the insets of Fig. 3) are caused by cross phase modulation (XPM) between the soliton and the coexisting CW component, which results in the same comb repetition rate with the soliton. By including the XPM effect from the single blue-detuned CW peak in the simulation, we achieve the XPM induced combs coexisting with the main soliton combs (see Supplementary Information Section S10), which agrees with the experimental results.

As for the long-term stability, soliton can keep stable for tens of minutes in the experiment. Due to the detuning variation resulting from the optical path length change in the large fiber cavity, the soliton repetition rate fluctuates in a range of 20 kHz for the 2-FSR microcombs, as shown in the inset of Fig. 3a (left panel). We believe passive and active control methods can be implemented to improve the long-term stability (see Supplementary Information Section S11).

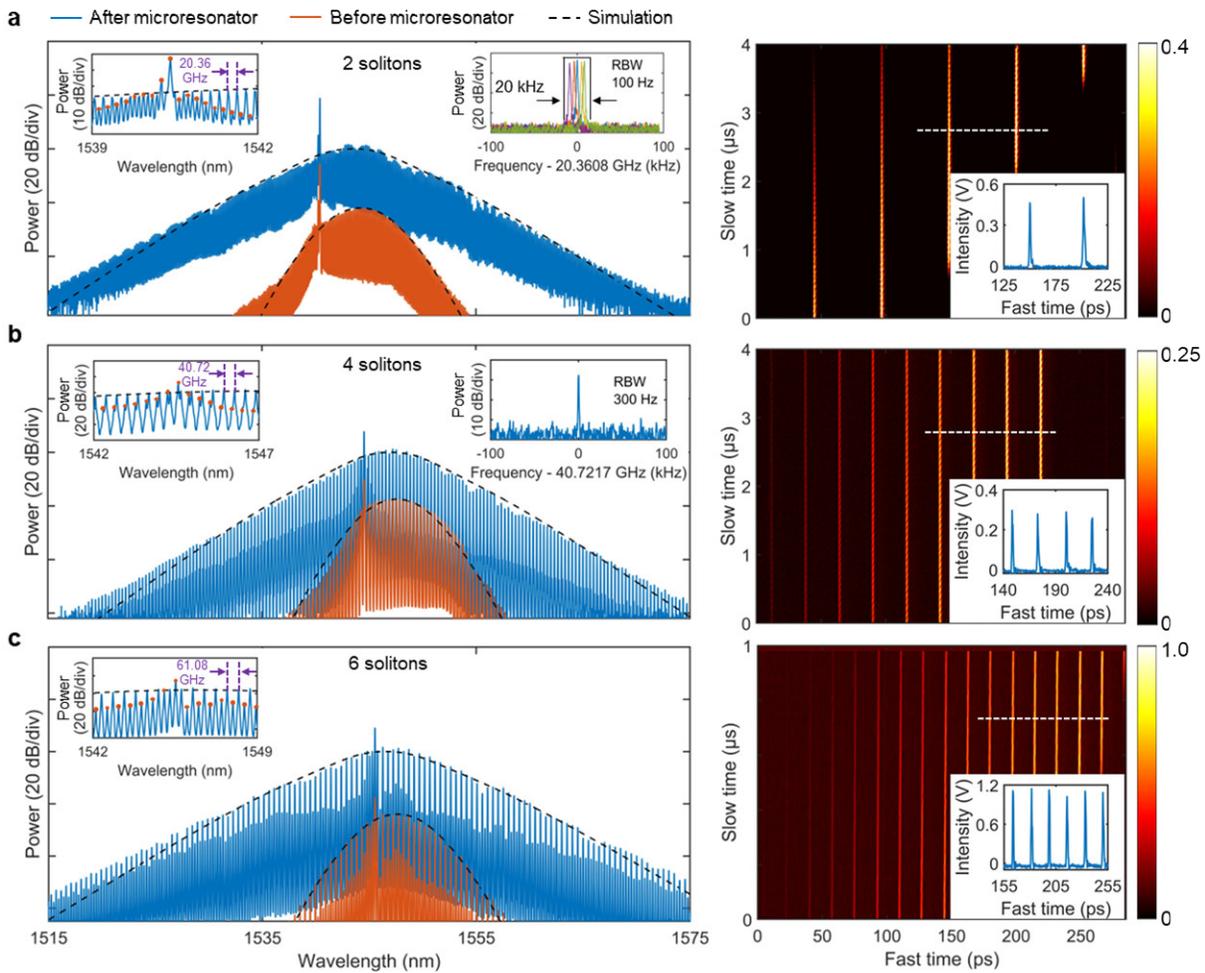

Fig. 3. Experimental generation of perfect soliton crystals. (a)-(c) soliton crystal with 2-FSR (a), 4-FSR (b) and 6-FSR (c). Left: optical spectrum. There is an offset between the spectra before and after the microresonator to clearly show the envelope of the XPM induced combs. Inset: enlarged spectrum and RF beat note of the soliton repetition rate. XPM comb peaks are indicated by orange solid circles. RF beat note of the soliton repetition rate with 60 GHz is not measured due to the lack of commercial electronics (see Methods). Right: real-time evolution of stable soliton recorded by time magnifier system. Inset: temporal trace calibrated by the responsivity curve. The white dashed line shows the time window of the calibrated trace.

**Dynamics of dissipative soliton formation and interaction.** Dissipative soliton in microresonator-filtered fiber lasers is fundamentally different from both the conventional mode-locked fiber lasers and externally driven microresonators. To gain more insight into the underlying physics, we study the soliton formation and interaction dynamics in both frequency domain and time domain. Since the pulse repetition rate is larger than 10 GHz, the traditional soliton dynamics monitoring by dispersive Fourier transform [59,60] is not feasible due to the insufficient time stretch limited by the pulse period. Instead, we use the powerful time magnifier system to study soliton dynamics in the presence of CW components induced by the phase shift at BPF edges.

Figures 4a-4c record the spectrum evolution from CW laser to soliton by manually and monotonously shortening the cavity length via the delay line. Six typical states are recorded when the delay line is paused at six positions. As shown in state I (Fig. 4a), a CW component lases first around the short-wavelength edge of the BPF (1542.4 nm), although it does not belong to the final soliton. In state II, another CW component stands out around the long-wavelength edge of the BPF (1549.7 nm), while primary comb lines are generated from non-degenerate FWM process. Thereafter sub-combs are generated near the primary combs (state III) through either degenerate or non-degenerate FWM process. In state IV, more combs start to lase due to the MI. Simultaneously, noisy pulses can be formed (see Supplementary Information Section S12), which is a precursor for mode locking. However, the pulses can not keep stable since they are in the unstable MI regime.

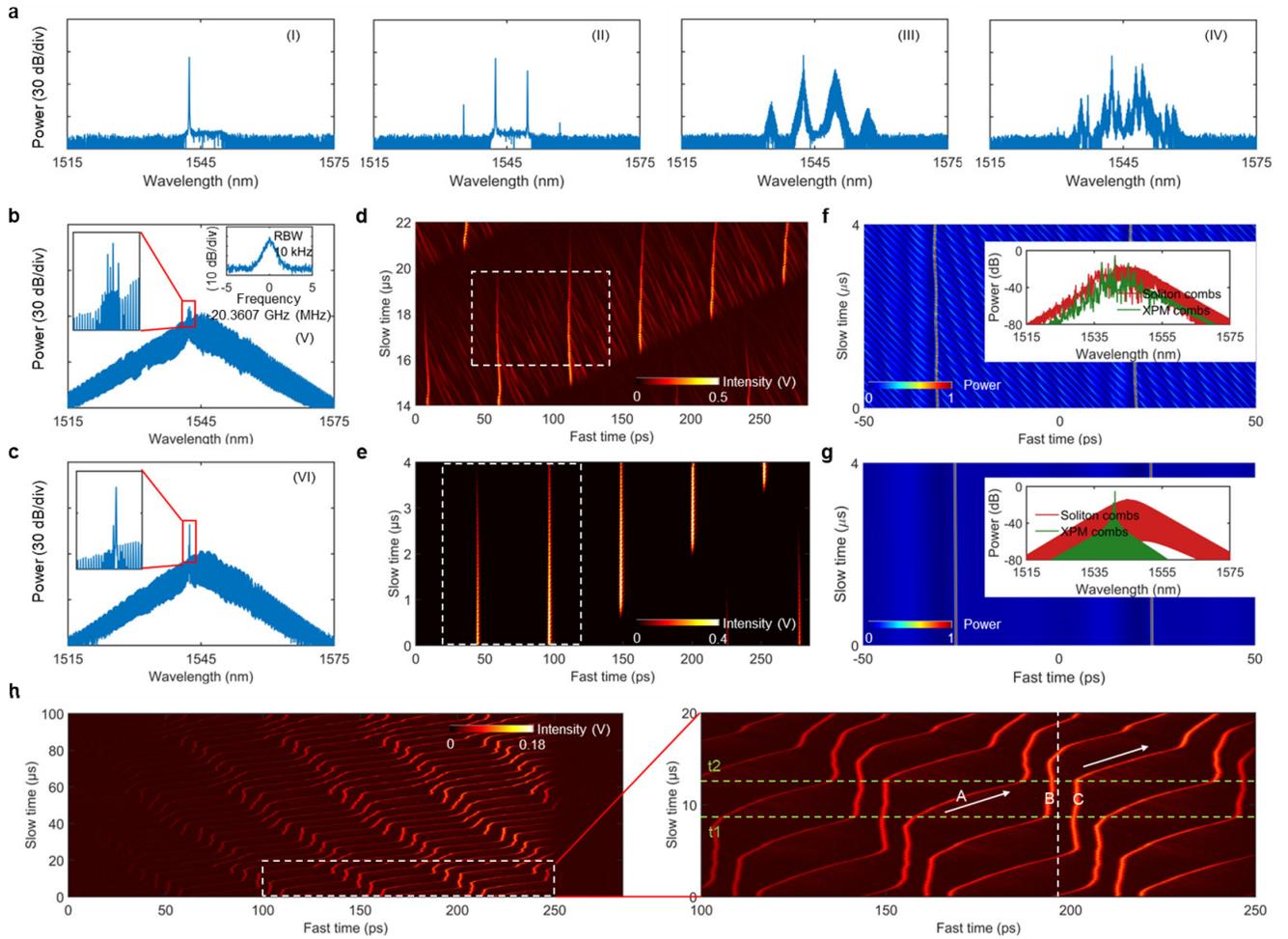

Fig. 4. Dynamics of dissipative soliton in microresonator-filtered fiber lasers. (a)-(c) Spectral evolution of different states when the detuning changes from the blue side to the red side. Insets in state V show the enlarged spectra and RF beat note of unstable repetition rate. Insets in state VI show the enlarged spectra. (d)(e) Real-time dynamics measured by time magnifier for (d) unstable mode-locking and (e) stable mode-locking. The white dashed boxes indicate the dynamics in a single roundtrip, which agree well with the simulated results. (f)(g) Simulated dynamics for (f) unstable mode-locking and (g) stable mode-locking. Insets are the simulated optical spectra. (h) Soliton Newton's cradle. The right panel is the zooned-in area of the left panel indicated by the white dashed box. The white dashed line in the right panel indicates the retarded time frame of soliton B. The arrowed lines indicate the soliton moving directions.

When the soliton comb lines are red detuned, the spectrum eventually evolves into a sech$^2$ profile (state V), with coexisting multi-frequency CW components (inset of Fig. 4b). Nonetheless, the comb intensity fluctuation and the broadband RF beat note (inset of Fig. 4b) of the repetition rate indicate the incoherent comb lines and unstable mode-locking. While in time domain (Fig. 4d), the solitons sit on a noisy background, interact and exchange energy with it, leading to fluctuating soliton temporal distances and unstable repetition rate. In the retarded time frame of soliton, the weak-intensity longer pulses in the background do not move at the same speed with the soliton. In fact, the weak-intensity longer pulses experience much smaller Kerr nonlinearity than the solitons so that they move faster than the solitons. The unstable mode-locking state is attributed to the strong perturbation from XPM effect between the coexisting CW components and soliton, which is well reproduced by simulation in Fig. 4f.

By further fine adjusting the delay line, the multiple CW components are suppressed and only single-frequency lasing coexists with the two solitons (state VI, Fig. 4c). In time domain (Fig. 4e), the solitons stably exist with a clean background and keep the temporal distance between each other. The narrowband RF beat note of the repetition rate indicates the stable mode-locking, as shown in the inset of Fig. 3a (left panel). The coexisting single CW peak can interact with soliton and generate additional combs through the XPM effect, which is verified by the inset of Fig. 3a (left panel) and the simulation results in Fig. 4g. However, the soliton stability is not affected due to the alleviated perturbation from the coexisting single CW peak and the strong SA effect from the large red detuning (Fig. 1c).

Besides, in state VI, we suddenly block the fiber laser cavity and in Fig. 4h record the interaction of 6 solitons circulating in the microresonator or more than 1400 solitons in the large fiber cavity. As shown in Fig. 4h in the retarded time frame of soliton B, at moment t1 soliton A moves towards a soliton molecule where soliton B and soliton C bind together and move at the same group velocity. As soliton A move closer, it starts to interact with the soliton molecule. At time t2, soliton A collides elastically with the soliton molecule and binds together with soliton B, forming a new soliton molecule. During the collision period, soliton B does not change the group velocity while soliton C gains the soliton A's group velocity and starts to move away from soliton B.

This collision scenario is similar to the Newton's cradle in classical mechanics [38–40], where the first ball moves in one direction and collide with another two balls remaining at rest with the same mass, leading to the energy and momentum exchange between the first and the third ball. As a consequence, the first ball stops moving and remains at rest together with the second ball, while the third ball starts to move at the same speed with the first ball before the collision. While in microresonator-filtered fiber lasers, solitons are identical with the same pulse duration and pulse intensity. In this dissipative system, the conservation of velocity difference between the single soliton and the soliton molecule before and after the collision, is mainly owed to the fixed stationary solutions governed by the system [37].

Two sets of this soliton Newton's cradle are found in the microresonator and they independently accomplish their corresponding elastic collision at the same time, mainly due to the large temporal distance and weak interaction. The elastic collision self-reproduces over time due to the cavity boundary conditions, which can be seen more clearly in the video in the Supplementary Information.

The soliton Newton's cradle can be achieved in a limited range of parameters in the microresonator-filtered fiber lasers, among many other possibilities for collision between three solitons [36]. In our case, the soliton molecule with non-zero binding energy is mainly attributed to the soliton interaction mediated by the coexisting CW component [53,55,61]. Compared to the single soliton, the soliton molecule with modulated spectrum has a different group velocity, leading to the walk-off at a speed of ~2000 m/s and eventually the collision with the single soliton. Benefitting from the fast interaction dynamics, the microresonator-filtered fiber lasers can act as a good test bench to study the soliton Newton's cradle in other fields.

In summary of the real-time dynamics, the coexisting CW components not only influence the soliton stability and the mode efficiency, but also induce interesting soliton interaction such as perfect soliton crystal and soliton Newton's cradle.

## Discussion
In general, dissipative soliton generation and interaction in microresonator-filtered fiber lasers are quite similar to that in the mode-locked fiber lasers. The microresonator-based SA and pulse shaper, however, differentiate the microresonator-filtered fiber lasers from conventional fiber lasers and provide unique property of microcombs with ultrahigh repetition rate. Compared with externally driven system, by introducing EDFAs and intracavity filters, the microresonator-filtered fiber lasers add more freedoms to adaptive control the soliton properties, such as self-starting, repetition rate [26] and spectral envelope. Besides, the intracavity spectral filtering provides a feasible method to generate bright broadband soliton with normal-GVD microresonators. However, every coin has two sides. The additional freedoms might cause the problem of soliton stability, especially the long-term stability, owing to the environmental perturbation to the long-length fiber cavity thus the detuning (see Supplementary Information Section

S11). Of course, this property in turn provides another route for optical sensing using the solitons in microresonator-filtered fiber lasers.

In conclusion, we comprehensively study the mode-locking in microresonator-filtered fiber lasers, including the principle, experimental guidelines and dynamics of soliton generation and interaction. Two roles of the microresonator for mode-locking, the SA and the nonlinearity booster, are identified in both theory and experiment, as well as the pulse shaping mechanism. Theoretical analysis shows that both GVDs of the microresonator and EDFA are important for the soliton mode-locking principle and properties. Chirped bright dissipative soliton with flat-top spectral envelope in normal-GVD microresonators is predicted, in contrast to the dark soliton for externally driven system. Experimental guidelines are given and soliton microcombs with 3-dB bandwidth of ~10 nm and repetition rates ranging from ~10 GHz to ~100 GHz are achieved. High soliton output power of ~12 mW with mode efficiency up to 90.7%, is also obtained. Soliton phase noise comparable to the on-chip platforms is demonstrated with -97 dBc/Hz, -117 dBc/Hz, and -137 dBc/Hz at 10 kHz, 100 kHz, and 1 MHz offset frequencies, respectively. Real time observations of soliton formation and interaction dynamics including the soliton Newton's cradle with coexisting CW components are demonstrated via an ultrahigh-speed time magnifier. Our study will benefit the design of the novel, high-efficiency and self-starting microcombs for real-world applications.

## Materials and Methods
### Experimental setup and measurement.

Our high-Q FFP microresonator is fabricated through three steps: (i) commercial HNLF (OFS 80412m2) is carefully cleaved and encapsuled in a ceramic fiber ferrule; (ii) both fiber ends are mechanically polished to sub-wavelength smoothness [62]; (iii) both fiber ends are coated with optical dielectric Bragg mirror (inset of Fig. 2a) with reflectivity over 99.5% from 1530 to 1570 nm. The FFP microresonator length is 10 mm, and the roundtrip length is 20 mm. The GVD of the FFP microresonator is $2.0 \pm 1.0$ ps/(nm·km) with a small dispersion slope of $0.019 \pm 0.004$ ps/(nm$^2$·km) for efficient nonlinear spectral broadening. The full width at half maximum (FWHM) linewidth of the microresonator is measured via frequency scanning calibrated by an unbalanced Mach–Zehnder interferometer (MZI) with a FSR of 1 MHz (see Supplementary Information Section S13). The transmission of the cold FFP microresonator is measured through Pound–Drever–Hall (PDH) locking technique. The microresonator is temperature controlled with a resolution of 10 mK.

As for the EDFA, the dispersion for the passive sing-mode fiber (~ 3.4 m) and the Erbium-doped fiber (~ 1 m) are -23 fs$^2$/mm and 59 fs$^2$/mm [63] at ~1550 nm, resulting in small, averaged dispersion of -4.4 fs$^2$/mm. In order to shorten the length of the large fiber cavity and ensures low coupling loss, free-space components are introduced to couple the light into and out from the FFP microresonator. We believe all-fiber integration is feasible due to the compatibility between the FFP microresonator and other fiber components. The home-built EDFA is not temperature controlled but is enclosed in a box made of acrylic sheet and is largely isolated from the lab environment, including the phonons and heat change through the air. The long-term temperature variation in the lab is $\pm 1$ °C. The optical isolator ensures the unidirectional lasing of the fiber cavity. A delay line is used to adjust the length-matching between the large fiber cavity and the FFP microresonator, as well as the relative position of the oscillating modes within the FFP microresonator.

The soliton phase noise is measured by injecting the combs into a fast photodetector (PD) (>10 GHz bandwidth), dividing the 10.18-GHz signal by 8 times, mixing the divided signal with a down-converter and detecting the down-converted signal with a phase noise analyzer (PNA).

A self-heterodyne frequency discriminator using a fiber-based unbalanced MZI and a balanced photodetector (BPD) is employed to measure the laser phase noise and fundamental linewidth. One arm of the unbalanced MZI is made of 250-m-long single mode fiber, while the other arm consists of an acousto-optic frequency shifter with frequency shift of 200 MHz and a polarization controller for high-voltage output. The FSR of the unbalanced MZI is 0.85 MHz. The two 50:50 outputs of the unbalanced MZI are connected to a BPD (PDB570C, Thorlabs) with a bandwidth of 400 MHz to reduce the impact of detector intensity fluctuations. The balanced output is then analyzed by a PNA. Fundamental linewidth is calculated from the white noise floor of the measured SSB frequency noise spectra. The minimum fundamental linewidth that can be measured by this frequency discriminator is below 10 mHz.

The RF beat notes of comb repetition rate in the insets of Figs. 2d and 2e are measured by injecting all the comb lines into an electro-optic intensity modulator modulated at 20 GHz. Each comb line will generate sidebands at 20 GHz and its harmonics. For the 2-FSR combs in Fig. 2d, the beat note of the comb repetition rate at 20.36 GHz is down-converted to the beat note between the 1$^{st}$-order sideband and the adjacent comb lines at 0.36 GHz. For the 4-FSR combs in Fig. 2e, the beat note of the comb repetition rate at 40.72 GHz is down-converted to the beat note between the 2$^{nd}$-order sideband and the adjacent comb lines or between the two 1$^{st}$-order sidebands from adjacent comb lines at 0.72 GHz. For the 6-FSR combs, the intensity of 3$^{rd}$-order sideband is too weak so that we can not measure the down-converted beat note with good signal-to-noise ratio. The down-converted beat note is detected by a fast PD and analyzed by an electrical spectrum analyzer (Agilent E4407B).

**Time magnifier setup.** The pump of the time magnifier is generated from a 50 MHz mode-locked fiber laser (Menlo C-fiber), which is chirped by around 400-m of Corning DCM-D-080-04 and amplified by a C-band EDFA. The signals under test are chirped by 200-m of Corning DCM-D-080-04 and the output dispersion is provided by 15 km of the same fiber. The final optical signal is pre-amplified by an L-band EDFA and then measured by a 38GHz-bandwidth photodetector (Newfocus 1474A) and subsequently digitized by a 25GHz real-time oscilloscope (Tektronix DPO72504D) with 100Gs/s sampling rate.

**Acknowledgments**
We thank Prof. David J. Moss and Dr. Hualong Bao for helpful discussions. We thank Prof. Juliet Gopinath for lending the high-speed oscilloscope. M.N., B.L., Y.X., J. Y. and S.W.H. acknowledge the support from the University of Colorado Boulder and National Science Foundation (OMA 2016244 and ECCS 2048202). K.J., S. Z. and Z.X. acknowledge the support by the National Key R&D Program of China (2019YFA0705000, 2017YFA0303700), Key R&D Program of Guangdong Province (2018B030329001), Leading-edge technology Program of Jiangsu Natural Science Foundation (BK20192001) and National Natural Science Foundation of China (51890861, 11690031, 11621091, 11627810, 11674169, 91950206).

**Author contributions**
M.N. and S.W.H. conceived the idea of the experiment. M.N. designed and performed the soliton generation and measurements with the help from Y.X.. B.L., M.N. and J.Y. designed and performed the time magnifier measurement. M.N. performed theoretical modeling and numerical simulation. K.J., S.Z. and Z.X. fabricated the FFP microresonator. M.N., B.L. and S.W.H. wrote the manuscript. S.W.H. led and supervised the project. All authors contributed to the analysis of the data and discussion and revision of the manuscript.

**Data availability.** The data that support the plots within this paper and the findings of this study are available from the corresponding authors on request.

**Competing interests**
The authors declare no competing interests.